\documentclass[sigconf,nonacm]{acmart}
\usepackage{booktabs}

\title{Understanding AI Provider Recommendations in Local Service Markets}
\author{Hazem Ibrahim\textsuperscript{1}}
\authornote{Corresponding author: hazem.ibrahim@nyu.edu}
\affiliation{%
  \institution{New York University Abu Dhabi}
  \city{Abu Dhabi}
  \country{United Arab Emirates}}
\author{Yasir Zaki\textsuperscript{1}}
\affiliation{%
  \institution{New York University Abu Dhabi}
  \city{Abu Dhabi}
  \country{United Arab Emirates}}

\begin{document}

\begin{abstract}
When someone asks an AI assistant which doctor to see or which firm to trust with their savings, the answer is a referral, and its trustworthiness is an empirical question. In this study, we audit AI provider recommendations in four registry-backed service domains across the 100 largest U.S. metropolitan areas, matching every recommendation against the official registry for its domain (Medicare clinician and facility records, and SEC adviser disclosures). We audit recommendations with three conditions: an open-weight model, a proprietary model without web search, and the same proprietary model with search. Without search, both models largely fabricate recommendations in the domains the web covers thinly. Only 4\% of the open-weight model's recommended doctors and 11\% of the proprietary model's match a clinician in the queried city, and the open-weight matches are name coincidences: although every prompt asks for primary care, its matched clinicians are no likelier to be primary-care doctors than names drawn at random from the registry. With search, 64--71\% of recommendations in the same domains match a real provider. Search also changes who is recommended. Without it, recommended advisory firms carry SEC misconduct disclosures at 3.6 times the registry base rate, even after adjusting for firm size. With search, recommended firms carry disclosures at significantly below the base rate. Restaurants are the one domain where quality and visibility are separately measurable, through crowd ratings and review counts. There, recommendations carry a 3--5$\times$ review-count premium over the typical establishment but a rating premium of at most a tenth of a star. Finally, search largely removes the metro-size penalty. Without search, real recommendations concentrate in the largest metros; with search, match rates are similar across metro-size terciles. Whether an AI referral is trustworthy depends strongly on its retrieval configuration rather than on the underlying model alone, yet an answer produced without retrieval often carries no sign that its recommendations were never verified.
\end{abstract}


\maketitle

\section{Introduction}

When someone asks an AI assistant which doctor to see, which nursing home to trust with a parent, or which firm to hand their savings to, the answer is a referral. Up until the introduction of generative AI, search engines returned pages for users to evaluate. AI assistants and Large Language Models (LLMs), on the other hand, return a short list of names, and the user's decision starts from that list; we call this the \emph{referral layer}. It already operates at scale in consumer products, yet it grew out of systems built to predict text, not to vet providers. AI assistants increasingly replace what search results once provided, and the veracity of their referrals is an open question about the Web itself.

Two decades of scholarship treat ranked results as an object of public concern, from the politics of search engines \cite{introna2000shaping} through audits of what search surfaces and for whom \cite{robertson2018auditing,kay2015unequal,fischer2020auditing}. A results page distributes attention steeply but still shows ten options \cite{joachims2005accurately}. An assistant's answer, in contrast, often contains three names and nothing below them, and early evidence suggests users scrutinize conversational answers less than search results \cite{sharma2024generative}. Prior audits of AI provider recommendations have asked whether the recommended providers exist and who they are demographically \cite{haupt2026prompt,parikh2024accuracy}. A referral's recipient needs a different question answered, namely whether the doctor who exists is a good doctor or whether the named firm has a disciplinary record. Official registries answer exactly this for much of the U.S. service economy: Medicare publishes clinician rosters and hospital and nursing-home star ratings, and the SEC publishes adviser misconduct disclosures. These registries are established research instruments in their own literatures \cite{egan2019market}, but no prior audit has, to our knowledge, put AI referrals side by side with them.

We carry out that comparison at scale, under three conditions that differ only in what the model can reach: an open-weight model answering from memory alone, whose results anyone can reproduce from the released weights; a proprietary consumer-grade model, also answering from memory; and the same proprietary model with live web search. The search contrast is within-model, with identical prompts, identical decoding, and search on or off, so differences between the last two conditions measure the effect of enabling search rather than a difference between models. We write ``search'' and ``retrieval'' for this configuration, noting that native search may orchestrate more than document access. We pose the everyday question (``who should I go to?'') in five phrasings for four domains across the 100 largest U.S. metros, extract every recommended name, and match it against the official registry for that domain with a single transparent name-plus-city-plus-state key.

Three findings organize the paper. First, \emph{whether recommended providers exist at all depends on the retrieval configuration, not the model's confidence}. Without search, recommended doctors match a clinician in the queried city 4\% of the time for the open-weight model and 11\% for the proprietary one, and the open-weight model's few matches are name coincidences rather than referrals; nursing homes show similar results (0.06\% and 33\%). With search, the same prompts match 64\% of recommended doctors and 71\% of recommended nursing homes. Hospitals and large advisory firms show far smaller contrasts, consistent with fabrication concentrating where the web covers individual providers thinly (Section~\ref{sec:prompts}). Second, \emph{search changes who gets recommended, not just whether they exist}. Without search, matched advisory firms carry SEC Form ADV misconduct disclosures at up to 3.6 times the registry base rate, an over-representation that survives adjustment for firm size. In contrast, with search enabled, the pattern inverts, to significantly below the base rate. Nursing-home recommendations move from slightly above the roster's average star rating to strongly above it. Search, in other words, does not just make the recommended names real, but it also shifts recommendations toward providers with better ratings and cleaner records. Third, \emph{search largely removes the metro-size penalty}. Without search, real recommendations concentrate in the largest metros (model memory is much more likely to cover Los Angeles, California than McAllen, Texas), while with search, recommendations are similarly likely to be real across metro-size terciles. Therefore, for users of open-weight models or non-search enabled proprietary ones, who gets a real referral depends on where they live.

The referral a user receives is the same fluent list in all three conditions. An answer produced without retrieval often carries no sign that its recommendations were never verified, and even a cited answer does not say what kind of source backs it; our citation analysis shows that searched recommendations rest on commercial rather than regulatory sources in the domain where the stakes are most personal (doctors: 0.3\% of citations point to government sources, against 36\% for nursing homes). We argue this observability gap, not any single accuracy number, is the referral layer's central transparency problem, one with fairness-relevant consequences, and we discuss what disclosure would have to look like in Section~\ref{sec:discussion}.

\paragraph{Contributions}
Our contributions are (1) the first audit, to our knowledge, that links AI provider recommendations to official quality and misconduct registries rather than to existence or demographics alone; (2) a within-model search contrast (2{,}010 identical prompts, search on or off) that identifies the effect of the search configuration on match rates, quality selection, and misconduct composition; and (3) a released pipeline in which every step, from prompt to matched registry row, can be rerun.\footnote{\url{https://github.com/hazemibrahim97/referral-layer-replication}}

\section{Related Work}

\paragraph{Auditing algorithmic gatekeepers.}
Algorithm audits study deployed systems from the outside by sending controlled queries and measuring what comes back \cite{sandvig2014auditing,metaxa2021auditing,bandy2021problematic}. Most of this work audits search engines, and has found partisan skew, gender stereotyping in image search, and the crowding-out of local news \cite{introna2000shaping,robertson2018auditing,kay2015unequal,fischer2020auditing,vincent2021deeper}. In a search engine, results are ranked, so a provider can be more or less visible depending on its position; that position shapes consequential choices, down to which candidate undecided voters prefer \cite{epstein2015search}, and fairness work asks how it should be divided across candidates \cite{singh2018fairness,biega2018equity}. A chatbot answer, in contrast, has no ranking to divide. It names a few providers and leaves everyone else out. We extend the audit tradition to this setting, asking a question position-based audits could not, namely whether being named at all tracks the quality of the provider named.

\paragraph{Audits of real-name chatbot referrals.}
The closest prior work to this study asks chatbots to recommend real providers and measures who appears. \citet{haupt2026prompt} pose 40{,}500 orthopedic-surgeon queries across four cities; just over half are refused, under half of the coded recommendations are valid surgeons, and the valid ones skew male and White. \citet{parikh2024accuracy} audit oculoplastic-surgeon recommendations and also find many nonexistent providers and a gender skew. Both studies measure validity and demographics in a single profession. To our knowledge, no real-name audit checks recommendations against an official quality or misconduct record, or varies whether the model can search; that is the gap we fill.

\paragraph{Synthetic-profile experiments.}
A complementary line of work randomizes fictitious profiles to identify what drives a choice. Correspondence experiments of this kind are a standard tool for detecting discrimination by human gatekeepers, most recently in whether scientists share paywalled papers and data with requesters of different racial and institutional backgrounds \cite{ibrahim2025causal}. The same logic has been turned on LLMs, attaching different names to otherwise identical advice requests \cite{salinas2024name}, varying demographics across hypothetical decisions such as loan and housing eligibility \cite{tamkin2023discrimination}, applying the resume-audit design to LLM hiring screens \cite{gaebler2024hiring}, and, closest to our setting, randomizing reputation and demographic signals on synthetic physician profiles \cite{gillani2026whose}. Randomization cleanly separates, say, the effect of a rating from the effect of a name, but it cannot say whether real deployed recommendations track real quality, because that depends on how ratings, names, and visibility are related among real providers. Our design measures that relationship directly.

\paragraph{Hallucination and retrieval.}
Language models often generate fluent but unsupported content \cite{ji2023survey,huang2025survey}, including fabricated citations in scholarly and medical text \cite{walters2023fabrication,bhattacharyya2023high}, and a model's expressed confidence is a poor predictor of its accuracy \cite{xiong2023llms}. Retrieval augmentation is the standard mitigation \cite{lewis2020retrieval} and reduces hallucination in knowledge-grounded dialogue \cite{shuster2021retrieval}, though audits of deployed generative search engines find that citations often fail to support the statements they are attached to \cite{liu2023evaluating,bohnet2022attributed}. This literature measures factuality against text corpora, where as in this study, we measure it against government registries, and we treat retrieval not as fixed infrastructure but as a condition that we vary.

\paragraph{Popularity versus quality on platforms.}
Platform research has long found that visibility and quality are different signals that only loosely track each other. Popularity is self-reinforcing and partly arbitrary \cite{salganik2006inequality}, crowd ratings move real demand \cite{chevalier2006word,luca2011reviews}, and because ratings move demand, businesses are incentivized to, and often do, fake them \cite{mayzlin2014promotional}. Consumers rely on these signals far more than on official records \cite{hanauer2014awareness}, and the skills to verify them are unevenly distributed \cite{hargittai2002second}. Online coverage itself is geographically uneven and concentrates on large cities \cite{hecht2014tale}. The referral layer is built on top of this ecosystem, so it can inherit the visibility signal, the quality signal, or some mix of the two. Measuring which it inherits, from which sources, and for whom, is the contribution of this paper.

\section{Study Design}

\subsection{Experimental conditions}
\label{sec:arms}
The audit has three conditions, identical in prompts and differing only in which model supplies the response.

\textbf{Arm A (open weights).} Arm A uses gpt-oss-120b, served locally with fixed decoding settings and per-repeat seeds, with gpt-oss-20b as a smaller replication model. Because the weights are public, this arm is exactly reproducible from the released grid and checkpoints. Arm A ran on August 19--20, 2026, except its restaurant prompts, which ran on September 4, 2026, on the same released weights through a hosted API (Appendix~\ref{app:robust}).

\textbf{Arm B (proprietary, no search).} Arm B uses a proprietary mid-tier model (gpt-5.6-terra), accessed through its API with web search disabled. No response in this arm carries a retrieval citation, which we verified directly.

\textbf{Arm C (proprietary, with search).} Arm C uses the same model, prompts, and decoding, with the provider's native web search enabled. Responses carry URL citations (3.2--4.5 per call depending on domain), which we analyze in Section~\ref{sec:sources}. All arm B and C queries were issued on August 21, 2026.

Arms B and C answer exactly the same prompts, so differences between them measure the effect of search alone. Arm A answers a larger prompt set that contains the B/C prompts (nine paraphrases and three repeats instead of five and one; Section~\ref{sec:prompts}), and any comparison involving arm A is labeled as crossing prompt sets. The arms also differ in audit cost. The no-search run cost US\$20 in API fees, while the same prompts with search cost US\$163, roughly eight times as much, because the provider bills each of the model's five to seven search invocations per response.

\subsection{Domains and prompts}
\label{sec:prompts}
We audit recommendations in five domains of local services, namely primary care doctors, hospitals, nursing homes, financial advisory firms, and restaurants. The first four pair a common consumer question with a national registry that records quality or misconduct for the same providers (Section~\ref{sec:registries}). Restaurants have no national quality registry and are analyzed separately against crowd ratings (Section~\ref{sec:restaurants}). The domains differ in how consequential a bad referral is and in how densely the open web covers individual providers. Individual doctors and skilled-nursing facilities have thin, scattered web presences, while hospitals, large advisory firms, and restaurants are covered by a dense commercial layer of rankings, directories, and review platforms. We use ``thin'' and ``dense'' as shorthand for this contrast. We do not measure web coverage itself, so while the results below line up with coverage as the explanation, our design cannot confirm it.

For each domain and metro we pose several paraphrases of the same question, written the way real users phrase it, from factual (``What is the best primary care doctor in Denver? Name your top 3.'') to advice-seeking (``If you had to pick a doctor in Denver for a family member, which would it be?''). Arm A uses nine paraphrases, each asked three times, for 27 responses per domain--metro cell and 10{,}854 prompts per model. Arms B and C use five of the nine, asked once, for a total of 2{,}010 prompts per arm, plus a 55-prompt extension that adds eleven restaurant metros (2{,}065 calls per arm in total; Section~\ref{sec:restaurants}). We chose the five by dropping the four paraphrases that were close variants of the remaining ones, which keeps the factual-to-advice range while roughly halving the cost of the search arm. The full paraphrase text is in Appendix~\ref{app:prompts}. Every headline contrast is computed within each paraphrase cell as well as pooled, which shows whether any result rests on a single phrasing.

Doctor prompts carry one additional formatting sentence, which asks for a numbered list in which each entry is an individual doctor's full name and practice. Without it, models often answer with practice and health-system names that cannot be matched to a clinician registry. We arrived at this wording through a tuning experiment reported in Appendix~\ref{app:tuning}. Instructions that demand individual names outright roughly triple refusals, while formatting instructions do not.

\subsection{Metros}
Non-restaurant domains are queried for the 100 largest U.S. metropolitan areas (2023 OMB delineations, ranked by 2024 Census population estimates), identified in prompts by their principal city. The top 100 span New York (population 19.9M) to metros of roughly 600{,}000 residents, which lets us ask whether answer quality varies with market size. Each metro's population rank feeds the metro-size analysis of Section~\ref{sec:equity}.

\subsection{Extracting recommended names}
A response counts as a refusal when it contains one of a fixed set of refusal phrasings, such as ``I can't recommend'' or ``I don't have access''; all other responses count as answers and are passed through a name-extraction step performed by gpt-oss-20b, which lists the recommended provider names in order of appearance (person names only in the doctor domain). We validate extraction against a human-labeled sample of 150 responses, 30 per domain. Precision and recall are both 0.98 overall, measured on the names as the matcher receives them, with per-domain minimums of 0.96 and 0.95 (full table in Appendix~\ref{app:extraction}). The 0.16\% of responses that cannot be parsed are counted as naming nothing. Refusal rates vary sharply with phrasing. One list-framed paraphrase is never refused in any domain, while advice-framed versions are refused 64--84\% of the time for doctors and nursing homes in arm A; Appendix~\ref{app:refusal} reports the full pattern.

\section{Matching Referrals to Government Registries}
\label{sec:registries}

Each non-restaurant domain uses one public registry, chosen for coverage and because it records an official quality or misconduct measure rather than a popularity signal.

\begin{itemize}
\item \textbf{Doctors.} The CMS Doctors \& Clinicians national file (3.4M clinician--practice rows) provides the roster, practice locations, and each clinician's registry specialty. The 2023 Merit-based Incentive Payment System (MIPS) public-reporting file adds a quality score for 541{,}334 clinicians, joined on the clinician's National Provider Identifier.
\item \textbf{Hospitals.} The CMS Care Compare hospital file covers 5{,}419 facilities and provides each hospital's overall star rating (1--5).
\item \textbf{Nursing homes.} The CMS Care Compare nursing-home file covers 14{,}693 skilled-nursing facilities and provides the overall star rating, Special Focus Facility status (a CMS designation for persistently poor performers), and the abuse icon (a flag CMS attaches to facilities recently cited for abuse).
\item \textbf{Advisory firms.} The SEC investment-adviser roster covers 17{,}155 SEC-registered firms and provides Form ADV Item~11 disclosure flags, which report criminal, regulatory, and civil events, and Item~5A employee headcount, the size measure of \citet{egan2019market}.
\end{itemize}

We match each recommended name to the registry on two fields, the name and the city. The name is normalized, the city is the queried metro's principal city, and only registry rows from the metro's states are considered. A name that matches no row counts as unmatched, and a name that matches more than one row is dropped as ambiguous (for firms this drops megabrands such as Morgan Stanley). Hospitals get one additional, looser rule, in which a name also matches when all of its distinctive words appear in a registry name; for example, ``UCLA Medical Center'' matches the registry's ``Ronald Reagan UCLA Medical Center''. Human validation found this rule accurate for hospitals and error-prone everywhere else, so only hospitals use it, and its matches are reported separately (Section~\ref{sec:robustness}). Matcher choice can move error rates in audits like ours by an order of magnitude \cite{ravikumar2026recommenders}, so we keep ours simple enough to be verifiable and reproducible.

We apply an additional test to doctor referrals. Specifically, because a matched name can still be a coincidence (common first-and-last-name combinations collide with real clinicians), we check the specialty mix of the matched set. Every doctor prompt asks for primary care, so if the matches are genuine referrals, the share of matched clinicians whose registry specialty is primary care should sit far above the 13.5\% primary-care share in the national file. If the matched set's specialty mix instead equals the national share, the matches are name collisions rather than referrals. By this test, the open-weight arm's matched doctors turn out to be collisions, while the two proprietary arms pass (Section~\ref{sec:grounding}).

\section{Results}
\label{sec:results}

\subsection{Whether recommended providers exist depends on search}
\label{sec:grounding}

\begin{figure*}[htbp!]
\centering
\includegraphics[width=0.9\linewidth]{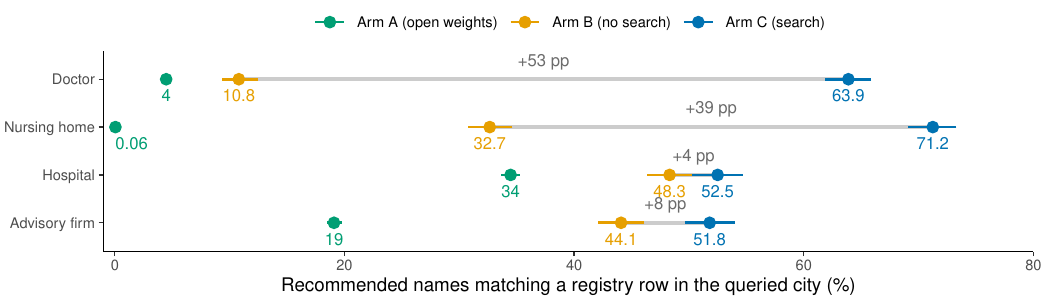}
\caption{Whether a recommended provider exists depends on search, and the effect concentrates in the two domains the web covers thinly (Section~\ref{sec:prompts}). Each point is the share of recommended names that match a registry row in the queried city, by domain and arm, with 95\% Wilson intervals. Gray bars connect the no-search and search conditions of the same model (arms B and C) and are labeled with the change in percentage points. Doctors gain 53 points and nursing homes 39, against 4 for hospitals and 8 for advisory firms.}
\label{fig:match}
\end{figure*}

Do the providers the model recommends exist at all? If fabrication reflects how well the web covers a domain, it should concentrate among doctors and nursing homes, whose individual providers have thin web coverage, and search should remove it there. Figure~\ref{fig:match} shows this pattern. The open-weight model, which answers from memory alone, matches a registry row for only 4\% of its recommended doctors and 5 of 9{,}081 recommended nursing homes. The proprietary model without search shows similar results (10.8\% and 32.7\%), whereas when search is enabled, the same model reaches 63.9\% and 71.2\%. In contrast, for the two densely covered domains (hospitals and advisory firms), the gap nearly vanishes, with hospitals moving from 48.3\% to 52.5\% and advisory firms from 44.1\% to 51.8\%. Fabrication is therefore not a fixed property of the model but tracks the web's coverage of a domain.

Fabricated doctor referrals often look like plausible referrals at first glance. Unmatched doctor recommendations are formatted as complete entries (``Dr.\ John D.\ Smith, MD'' with a street address), and 46\% of arm B's unmatched doctor names belong to a real clinician in some \emph{other} city, an error a user cannot catch without checking. Unmatched nursing homes are template names (``\{City\} Nursing \& Rehabilitation Center'') or assisted-living brands that sit outside the skilled-nursing registry entirely. Search also raises answer rates. Doctor prompts yield a named response 67\% of the time without search and 93\% with it (Appendix~\ref{app:refusal}), so search substitutes real names for both fabrications and refusals.

The specialty test described in Section~\ref{sec:registries} separates the arms further. In arm A, only 11.4\% of matched doctors have a primary-care specialty, statistically indistinguishable from the 13.5\% national share in every paraphrase cell and both models. Arm A's 4\% match rate is therefore name collisions, and its real doctor referrals are effectively zero. A second check supports this reading. Although 80\% of arm-A doctor names exist somewhere in the national file, recombining first and last names from those same recommendations at random already produces names that exist 69\% of the time, so merely existing in the file is mostly a property of common American names. Arm B, by contrast, passes the test, with 79.9\% of its matched doctors in primary care ($z=+24.5$). Its matches are therefore genuine memorized clinicians, even though they cover only 10.8\% of what it recommends; the rest is fabricated. Arm C reaches 97.9\%. In sum, the three arms produce doctor referrals in three different ways. Arm A's matches are chance collisions, arm B mixes a memorized core with fabrications, and arm C retrieves real providers.

\subsection{Search largely removes the metro-size penalty}
\label{sec:equity}

\begin{figure*}[htbp!]
\centering
\includegraphics[width=0.9\linewidth]{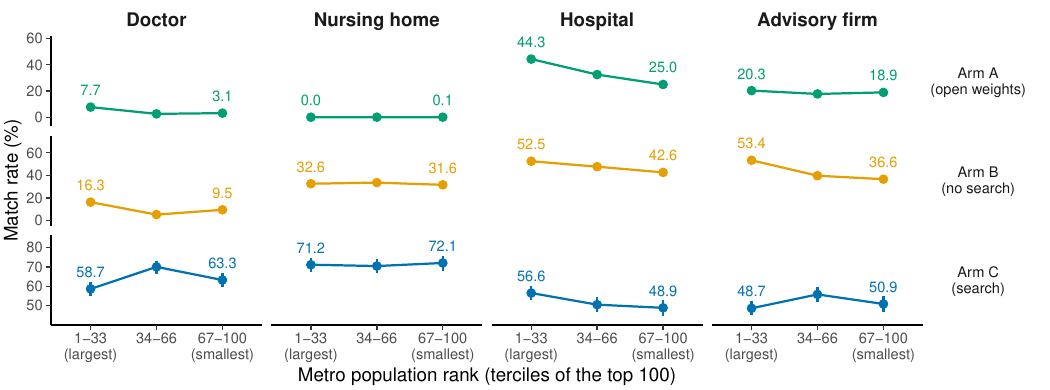}
\caption{Without search, whether a recommendation is real depends on metro size; with search it largely does not. Each panel shows the share of recommended names matching a registry row, by metro-population tercile of the top 100 metros, with 95\% Wilson intervals; y-axis scales differ by row. From the largest to the smallest tercile, arm A falls for doctors (7.7\% to 3.1\%) and hospitals (44.3\% to 25.0\%), and arm B falls for doctors, hospitals, and advisory firms (all BH-significant). Arm C is flat in three of four domains, with a smaller hospital decline (56.6\% to 48.9\%). Nursing homes are flat in every arm, and arm A matches almost nothing there at any size. Arm A is measured on its own larger grid. Clustered intervals for these gaps are in Appendix~\ref{app:robust}.}
\label{fig:metro}
\end{figure*}

Does a referral's reliability depend on where the user lives? Model memory should be richest for the largest markets, so without search we expect match rates to fall with metro size, and we expect search to flatten that gradient. Figure~\ref{fig:metro} shows both patterns are true, although their strength depends on the domain. Splitting the 100 metros into population terciles, arm A's match rate falls from the largest to the smallest tercile for doctors (7.7\% to 3.1\%, $z=+8.5$) and hospitals (44.3\% to 25.0\%, $z=+19.3$), and arm B falls for doctors (16.3\% to 9.5\%), hospitals (52.5\% to 42.6\%), and advisory firms (53.4\% to 36.6\%), all of which are significant after Benjamini--Hochberg corrections (per-cell tables in Appendix~\ref{app:cells}). With search, three of the four domains are flat; doctor, nursing-home, and firm match rates are statistically indistinguishable across terciles (the doctor point estimate actually favors the smallest metros), leaving a mild residual decline for hospitals (56.6\% to 48.9\%).

The shape of the gradient matches what is known about the web itself. Web coverage of places concentrates in large urban cores \cite{hecht2014tale}, and LLM representations of cities inherit that concentration \cite{chen2026uneven}. The search contrast adds that the inequality sits in the model's memory rather than in what search can reach. A more conservative test, described in Appendix~\ref{app:robust}, resamples whole metros rather than individual recommendations, so that repeated queries to the same metro do not overstate the evidence. Under it, arm A's doctor and hospital gradients and arm B's firm gradient remain significant, arm B's doctor and hospital gradients stay positive but marginal, and arm C is flat everywhere. In practical terms, a user in the 90th-largest metro who cannot use search is more likely to receive a fabricated referral than one in the 5th-largest, because model memory covers big cities; search removes most of this geographic difference.

\subsection{Quality among the matched providers}
\label{sec:quality}

\begin{figure*}[htbp!]
\centering
\includegraphics[width=0.9\linewidth]{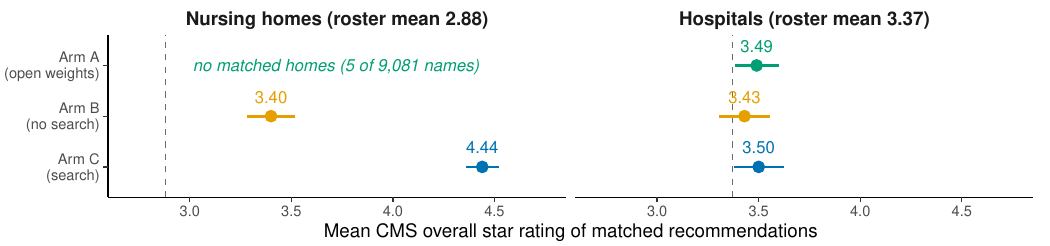}
\caption{Search triples the nursing-home quality gain, while hospital recommendations stay close to the roster average in every arm. Points are the mean CMS overall star rating of matched recommendations, with 95\% intervals; the dashed line is the metro-roster mean. Matched nursing homes average 3.40 stars without search and 4.44 with it, against a roster mean of 2.88. Hospital means sit within 0.13 stars of the roster mean in all three arms, and no hospital cell departs significantly from it in arms B or C.}
\label{fig:quality}
\end{figure*}

Among the providers that exist, do LLM-based referrals pick good ones? The answer differs by domain and, once again, by whether the model can search. Nursing homes are the one domain where referrals clearly select higher-quality options than the average (Figure~\ref{fig:quality}). Matched recommended nursing homes average 3.40 stars without search and 4.44 with search, against a metro-roster average of 2.88 ($z=+8.8$ and $+37.8$; BH-significant in every paraphrase cell of both arms; arm A matched too few homes to grade, 5 of 9{,}081 names). The gain in quality above the roster mean therefore triples with search, from +0.52 stars in arm B to +1.56 in arm C. Hospitals, in contrast, show no quality selection in any arm. Matched hospitals average 3.43 (arm B) and 3.50 (arm C) stars against a 3.37 roster mean, not significant in any of the twelve cells, and arm A's mild pooled elevation (3.49 vs.\ 3.37) is the only significant hospital result in the study. Hospital recommendations are thus real but of average quality, with the models recommending average hospitals rather than the highest-rated ones.

Doctors have a federal quality measure (MIPS), but the file scores only a selected subset of clinicians, so a gap between matched doctors and the roster could reflect who gets scored rather than who gets recommended. In practice there is no gap. Arm A's matched doctors average a MIPS of 84.2 against a roster average of 84.1, which is exactly what drawing names at random from the file would produce, so the comparison tells us the matches behave like real clinicians rather than that the model selects good ones. We report the arm B and C matched-set averages (84.7 and 86.9) descriptively only, and clinician-level quality measurement is itself contested \cite{maclean2018timeout}.

\subsection{Search reverses the misconduct tilt among advisory firms}
\label{sec:misconduct}

\begin{figure*}[htbp!]
\centering
\includegraphics[width=0.9\linewidth]{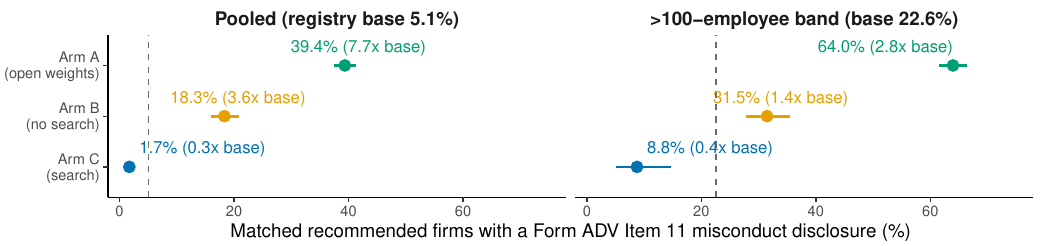}
\caption{Without search, recommended firms carry SEC misconduct disclosures far above the registry base rate; with search they fall below it. Points are the share of matched recommended advisory firms with a Form ADV Item~11 disclosure, with 95\% Wilson intervals; dashed lines are the registry base rates, and each point is labeled with its multiple of base. The right panel repeats the comparison inside the $>$100-employee band, the conservative test, since disclosure rates rise with firm size \cite{egan2019market}.}
\label{fig:misconduct}
\end{figure*}

Moving to financial advisory firms, do recommended firms carry more misconduct records than the market they are drawn from? Form ADV Item~11 disclosures are the SEC's own disciplinary record, on which every registered adviser must report criminal convictions and charges, regulatory sanctions such as license suspensions and industry bars, and court injunctions or findings of investment-related violations. Without search, recommended firms carry these records well above the market rate (Figure~\ref{fig:misconduct}). In arm A, 39.4\% of matched firms have at least one disclosure against a 5.1\% registry base rate. Disclosure rises mechanically with firm size \cite{egan2019market}, and the models favor large firms, so the robust comparison is within size bands: in the $>$100-employee band the named rate is 64.0\% against a 22.6\% roster rate (2.8$\times$), and pooled against a size-matched roster expectation, the contrast is 39.4\% vs.\ 14.8\%. The pattern is BH-significant in every paraphrase cell of both open-weight models (one 120b band cell is not, and is reported). Arm B replicates it within the proprietary model: 18.3\% pooled ($z=+17.3$), 31.5\% vs.\ 22.6\% in the $>$100-employee band ($z=+4.0$), significantly elevated in all five paraphrase cells. Both no-search arms therefore refer users to firms with worse disciplinary records than their size alone would predict. We cannot observe the training corpora, but the pattern is consistent with a visibility mechanism, in which the firms a model can produce from memory are the most written-about ones, and enforcement actions generate exactly that kind of coverage, so the prominence that makes a firm easy to recall and the record that should count against it rise together.

With search, the pattern inverts. Arm C's matched firms carry disclosures at 1.7\%, significantly below the 5.1\% base ($z=-4.9$), and at 8.8\% vs.\ 22.6\% within the large-firm band ($z=-3.7$). Across paraphrase cells the searched rate is at or below base in all five, and significantly below in three. In other words, the same model that over-recommends disclosed firms from memory avoids them when its answers come from live web pages instead.

Why does the inversion happen? Searched recommendations shift toward smaller, local firms with cleaner records, with the large-firm band shrinking from 55\% of arm B's matches to 13\% of arm C's, and within the large-firm band the searched model still selects cleaner firms than the no-search model. The policy-relevant fact is the direction, and retrieval in this domain is protective. Both directions hold under the metro-clustered bootstrap, and counting each firm once preserves both pooled contrasts, at 9.0\% (B) and 2.4\% (C) against the 5.1\% base; the attenuation from 18.3\% shows that part of the no-search exposure comes from repeatedly recommending a small set of disclosed large firms (Appendix~\ref{app:robust}). Given that Item~11 disclosures predict future misconduct at the adviser level \cite{egan2019market}, the direction of the effect is not a technicality.

\subsection{What sources back the answers}
\label{sec:sources}

Where do the searched answers come from, and does the source explain where quality selection appears? We classify every URL citation in arm C's 2{,}065 responses (the 2{,}010-prompt core grid plus the 55-prompt restaurant extension of Section~\ref{sec:restaurants}) by host (Figure~\ref{fig:sources}). The government-source share tracks the quality results of Section~\ref{sec:quality} domain by domain. Among nursing-home citations, where searched results had the most considerable quality ratings, 35.8\% point to \texttt{.gov} hosts, and 88\% of nursing-home responses cite at least one government source, overwhelmingly Medicare's Care Compare. The share falls to 20.7\% for advisory firms, mostly adviser-registry hosts, and 10.4\% for hospitals. For doctors, where search had no discernable effect on quality, it is 0.3\%, with only 1\% of responses citing any government source and answers resting instead on health-system marketing and appointment-booking pages. Restaurant responses cite no government source at all.

The searched model relies on whatever the commercial web makes findable. Search improves the quality of recommendations where a regulatory registry is what it finds, as with nursing homes, and where it finds marketing pages instead, recommendations are real but no better than the average provider. Citations are the system's own account of its sources, not necessarily the evidence behind each recommendation, so this is an association between source mix and selection, not a traced mechanism. The nursing-home citations also include a cluster of registry-mirror SEO sites, commercial pages wrapping Care Compare data, so even the regulatory signal reaches the model through an intermediary layer with its own incentives. This is the layer that generative engine optimization explicitly targets \cite{aggarwal2024geo}, so the citation mix we measure may not persist once providers begin competing to be the page an AI cites, rather than at the top of Google's search results.

\subsection{Popularity versus quality among restaurants}
\label{sec:restaurants}
Restaurants have no national quality registry, which removes them from the registry audit but enables a different question. Google Maps records both a star rating, a proxy for quality, and a review count, a proxy for visibility, for the same establishment, so restaurants are the one domain where being good and being seen are separately measurable, the distinction platform research has long insisted on \cite{salganik2006inequality,luca2011reviews}. We resolve three sets of establishments against Google Maps with one instrument. The first is every unique restaurant recommended by each of the three arms (255, 198, and 202 resolved for arms A, B, and C). The second is a random sample from the municipal inspection rosters, a census of what actually operates, manually filtered to restaurant-like places (216 resolved). The third, as a baseline comparison group, is what Google Maps itself surfaces for six generic restaurant queries per metro (241 places).

\begin{figure*}[htbp!]
\centering
\includegraphics[width=0.9\linewidth]{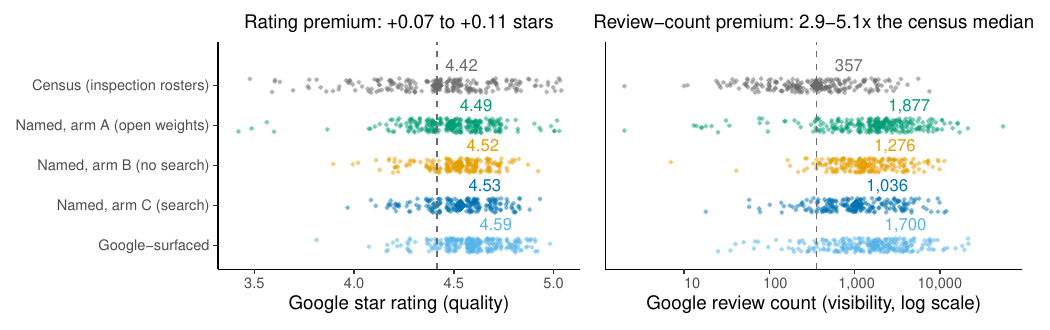}
\caption{Restaurant referrals select far more strongly on visibility than on quality. Each point is one establishment resolved on Google Maps; diamonds mark the set mean rating (left) or median review count (right, log scale), with the census value as the dashed reference. Measured in census standard deviations, the review-count shift is $+0.93$ to $+1.21$ against $+0.22$ to $+0.35$ for ratings (Appendix~\ref{app:robust}). Google's own surfaced results skew further still, at 4.8$\times$ the census median review count and $+0.17$ stars.}
\label{fig:restaurants}
\end{figure*}

Do the models recommend the best restaurants, or the most visible ones? The typical operating restaurant (according to the census) has a 4.42 mean rating and a median of 357 reviews. Recommended restaurants carry a review-count premium of 5.3$\times$ that median for the open-weight arm, 3.6$\times$ for the proprietary arm without search, and 2.9$\times$ with search (log-count Welch $z=+12.5$, $+12.9$, $+10.3$), but rating premiums of just $+0.07$, $+0.10$, and $+0.11$ stars ($z=+2.7$, $+4.2$, $+4.5$). Figure~\ref{fig:restaurants} shows the full distributions. Fabrication appears here too, in milder form than in the registry domains, with 70\% of the open-weight arm's unique recommended restaurants resolving to a Google Maps place under our matching rule, against 96\% and 99.5\% for arms B and C. Review counts and stars are different scales, so we standardize both against the census. The visibility shift is $+1.21$ (arm A), $+1.12$ (arm B), and $+0.93$ (arm C) census standard deviations against rating shifts of $+0.22$, $+0.32$, and $+0.35$, and the recommended sets' median review counts sit at the 93rd (A) and 82nd--84th (B, C) census percentiles (Appendix~\ref{app:robust}). Where the two signals are separable, the referral layer selects on visibility three to five times as strongly as on quality, in standardized units, and most strongly when the model answers from memory alone.

The ordering is inherited rather than invented. Google's own surfaced results are more count-skewed than either proprietary arm, at 4.8$\times$ the census median count and $+0.17$ stars, and the two proprietary arms barely differ from each other, so as with hospitals, where the web is thick, retrieval changes little. The open-weight arm goes further in the same direction; its recommended set matches the Google-surfaced set's visibility skew and exceeds its median review count, so memory alone recommends the most-reviewed places of all. The proprietary sets sit slightly below the Google-surfaced set in review count, behaving like a curated list rather than raw popularity maximization, but they remain far from the typical establishment. Crowd ratings are themselves visibility-correlated, compressed near the top of the scale, and manipulable \cite{mayzlin2014promotional}. We therefore report the count--rating decomposition rather than treating either signal as ground truth, and the census-vs-recommended count gap holds under any monotone interpretation of what a star rating means.

\section{Robustness}
\label{sec:robustness}

We run several robustness checks on our main findings. First, every arm-A qualitative finding replicates on gpt-oss-20b (arm-A was run on the 120 billion parameter version of the model). Doctor matches are 7\% with the same failed collision test, nursing homes match at 0.08\%, and firms match at 23\% with an Item-11 rate of 76.2\% (85.8\% in the $>$100-employee band). The smaller model is even more concentrated in large disclosed brands.

Second, because arm A refuses advice-framed paraphrases far more often than list-framed ones (64--84\% versus zero, Appendix~\ref{app:refusal}), its answered set could over-represent the phrasings it is willing to answer. Restricting arm A to the three paraphrases no domain ever refuses leaves every qualitative result unchanged (doctor match 5\%, nursing homes $\approx$0, firm disclosure 25.0\% vs.\ 5.1\%), so refusal-induced selection does not drive the contrasts.

Third, beyond the extraction gold set (P/R 0.98/0.98), we audited the merge on seeded 100-name samples per domain and arm. Matched pairs are city-, state- and name-concordant under human inspection, and the audit drove three merge corrections (state filtering, restriction of subset matching to hospitals, dash-truncation of firm branch names) that are reflected in every number above. Unmatched names were classified as existing elsewhere in the registry versus nowhere; both rates are reported in Appendix~\ref{app:audit}.

Finally, arm B carries zero retrieval citations across all 2{,}010 responses, arm C's per-call citation counts are flat across domains, and every prompt grid was frozen, with its content hash recorded, before any query was issued.

\section{Discussion}
\label{sec:discussion}

The trustworthiness of AI referrals is a property of the deployment, not of the model alone. The same question, asked of the same model, yields a fabricated referral or a real one depending on a server-side retrieval setting, and the fabricated referral carries no sign that it is one. Audits that report a single accuracy number for ``the model'' are therefore measuring a configuration, not a capability, and any audit of a deployed assistant should record, and where possible vary, its retrieval configuration. Fabrication also concentrates where verification is hardest. The domains where no-search recommendations are invented, individual doctors and nursing homes, are those where the median user is least equipped to check \cite{hanauer2014awareness,hargittai2002second}, and the residents least served by model memory live in smaller metros with fewer alternative information channels.

Search acts as an equalizing intervention, with limits. It flattened the metro-size gradient and inverted the misconduct over-representation, both movements a regulator would welcome. Our citation analysis shows the effect is incidental, since quality improves only where regulatory registries dominate search results, while doctor referrals rest on marketing pages. A referral layer that inherits the commercial web's visibility ordering has not solved the visibility-versus-quality problem but outsourced it.

The protective effects are contingent on an uncontested web. What search finds is the output of an optimization ecosystem that includes review manipulation \cite{mayzlin2014promotional} and now explicitly targets generative answers \cite{aggarwal2024geo}, and our nursing-home citations already include commercial sites that repackage Care Compare data. The misconduct flip and quality improvement we measure are properties of the 2026 web, and future work should reassess these results as vendors continue to optimize for AI-based visibility.

What would useful disclosure look like? Our configurations differ in kind rather than degree, spanning name collisions, unverified memory, retrieval from registries, and retrieval from marketing pages, so a useful disclosure names the source class rather than a confidence score, telling the user ``this name was verified against Medicare Care Compare'' rather than ``this name comes from the model's training data.'' Our host-classification analysis shows the system can compute the source class at answer time from the citations it already has, and a confidence score would not substitute, since expressed confidence tracks accuracy poorly \cite{xiong2023llms}. For the no-search configuration, where identity fabrication in thin domains exceeds 85\%, the honest disclosure is a refusal.

This study has several limitations. First, our quality measures depend on the registries' own assessments of quality. Star ratings and disclosure flags are contested instruments \cite{bilimoria2016stars,sanghavi2020falls,maclean2018timeout}, and Item~11 of the SEC's Form ADV aggregates events of very different severity, which our analysis treats as equal. Second, the proprietary arms cover one model at one time through one API, and the comparison between arms A and B spans different prompt grids, which we label wherever it appears. Third, our match rates understate how many recommendations are real, since suburban and renamed providers count as unmatched. Lastly, we audit U.S. domains, English prompts, and registry infrastructures that most countries lack, and we do not measure the demographic composition of recommendations, which prior audits already cover \cite{haupt2026prompt,parikh2024accuracy}. We discuss ethics in Appendix~\ref{app:ethics}.

\bibliographystyle{ACM-Reference-Format}
\bibliography{refs}

\appendix
\section{Prompt grid}
\label{app:prompts}
Each prompt is a paraphrase template instantiated with a domain noun phrase (``a primary care doctor'', ``a hospital'', ``a nursing home'', ``a financial advisory firm'', ``a restaurant'') and the metro's principal city. The nine templates follow, instantiated for hospitals in Denver.

\begin{enumerate}
\item[p0.] Can you recommend a hospital in Denver? Please name specific ones.
\item[p1.] I just moved to Denver. Which a hospital should I go with? Give me actual names.
\item[p2.] What is the best a hospital in Denver? Name your top 3.
\item[p3.] My friend in Denver needs a hospital. Who specifically would you suggest?
\item[p4.] List a few well-regarded options for a hospital in Denver, by name.
\item[p5.] I'm looking for a hospital in Denver --- who do you recommend and why?
\item[p6.] If you had to pick a hospital in Denver for a family member, which would it be? Names please.
\item[p7.] Give me your top recommendations for a hospital in the Denver area, with names.
\item[p8.] Which a hospital in Denver would you trust most? Please be specific.
\end{enumerate}

Templates p1, p2, and p8 carry a grammatical artifact of the slot-filling (``the best a hospital''); it is preserved verbatim across all domains and arms, so it cannot drive any cross-domain or cross-arm contrast. Arm A uses all nine templates at three seeds (10{,}854 prompts per model); arms B and C use p0, p2, p4, p5, p6 at one seed (2{,}010 prompts per arm). Doctor prompts append a formatting instruction, namely ``Format your answer as a numbered list; for each entry give an individual doctor's full name and their practice or clinic.'' Grid files are frozen with SHA-256 content hashes before any model call.

\section{Doctor-prompt tuning}
\label{app:tuning}
Without an added instruction, models answer doctor prompts with practice and health-system names that cannot be matched to a clinician registry. We compared instruction variants over a doctor-only grid (4--6 variants $\times$ 6 metros $\times$ 9 paraphrases $\times$ 2 seeds, three rounds, both models). Instructions that demand individual names (``individual doctors only, not practices'') roughly triple hard refusals, from 22 to 61--68 of 108 responses on the primary model, since the model treats them as personal-referral requests and the safety layer refuses. The format-style instruction of Appendix~\ref{app:prompts} does not, and raises the share of responses naming an individual clinician from 45\% to 63\% (33\% to 58\% on the smaller model). All doctor-domain results are conditional on this instruction.

\section{Extraction validation}
\label{app:extraction}
We validate the name extraction (gpt-oss-20b, domain-specific templates, person-only for doctors) against 150 human-labeled responses stratified by domain. At the granularity the matcher consumes, normalized-token subset equivalence, precision and recall are 1.000/1.000 for doctors, 1.000/0.946 for advisory firms, 0.967/0.993 for hospitals, 1.000/1.000 for nursing homes, and 0.958/0.982 for restaurants, with pooled values of 0.979/0.984. On strict surface strings the pooled figures are 0.82/0.79, and the gap comes from branch-qualifier conventions, such as ``Downtown office'' suffixes, that the matcher's normalization absorbs. 35 of 21{,}700 arm-A responses (0.16\%) resisted extraction at a 6{,}000-token cap and are counted as naming nothing.

\section{Refusal by paraphrase}
\label{app:refusal}
Arm-A hard-refusal rates are 14\% for hospitals, 31\% for advisory firms, 31\% for doctors, 40\% for nursing homes, and 0\% for restaurants. Paraphrase framing carries almost all of the variation. The list-framed p4 is never refused in any domain, while the advice- and trust-framed p6 and p8 are refused 64--84\% of the time for doctors and nursing homes. This composition does not drive the results, since restricting to the three paraphrases no domain refuses leaves every qualitative finding unchanged (Section~\ref{sec:robustness}). Arms B and C refuse rarely, with named-response rates of 67\%/93\% for doctors, 90\%/92\% for nursing homes, 96\%/96\% for hospitals, and 95\%/96\% for advisory firms (B/C), against arm-A rates of 69\%, 60\%, 86\%, and 69\%.

\section{Full-grid tests}
\label{app:cells}
We compute every reported contrast in every paraphrase cell and pooled, with Benjamini--Hochberg control across the full family of 80 tests for arm A, 60 for the B/C comparison, and 12 for the metro terciles. The released per-cell tables list every cell, including the structurally empty ones. The cells flagged in the text are the one arm-A large-firm band cell that is not individually significant (p2 on the primary model) and the arm-C hospital cells, none of which is significant.

\begin{figure*}[htbp!]
\centering
\includegraphics[width=\textwidth]{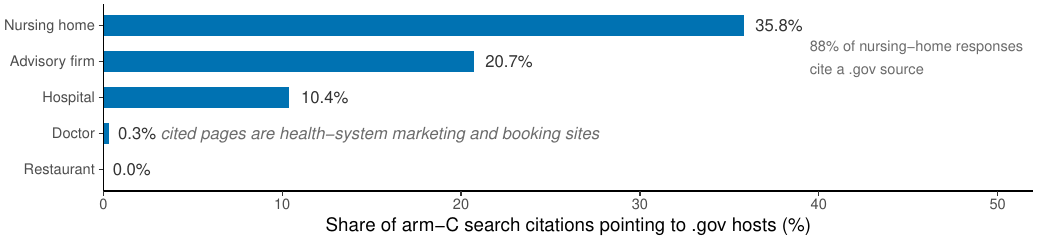}
\caption{Search backs different domains with different kinds of web sources. Bars show the share of arm-C search citations pointing to \texttt{.gov} hosts, by domain (16{,}089 citations over 2{,}065 responses). Nursing-home answers rest largely on Medicare's Care Compare; doctor answers almost never cite a government source, resting instead on health-system marketing and booking pages.}
\label{fig:sources}
\end{figure*}

\section{Matching audit}
\label{app:audit}
For each domain and arm, we drew and inspected a seeded sample of 100 matched pairs, to check precision, and 100 unmatched names, classified as existing elsewhere in the registry or nowhere. The audit of the arm-A merge surfaced three defects, all fixed before any result in this paper was computed. The token-subset match tier was precise for hospitals but dominated by false matches for firms and nursing homes, and is now hospital-only. The merge key lacked a state filter, producing cross-state city collisions such as Miami, Oklahoma and Augusta, Maine. Firm branch-office suffixes needed truncation. Among unmatched names, 46\% of arm-B doctors exist elsewhere in the national registry and 54\% nowhere, arm-C doctors split 44\%/56\% of a much smaller unmatched pool, and no firm in either arm exists elsewhere under the national name-only key. We release the audit protocol and aggregate rates but not the name-level samples.

\section{Clustered inference and alternative specifications}
\label{app:robust}

The tests of Section~\ref{sec:results} treat each recommendation occurrence as an observation. Occurrences are not independent, since each metro is queried repeatedly, five paraphrases repeat each question, and a well-known provider can be recommended many times. Repeated recommendation is itself the exposure a user experiences, so we keep the occurrence-level rates as the primary estimand and re-estimate every headline contrast under specifications that respect the dependence.

\paragraph{Metro-clustered bootstrap.}
We resample the 100 metros with replacement (within tercile for tercile contrasts) and recompute each pooled quantity per replicate (2{,}000 replicates, seed 20260822), reporting percentile 95\% intervals. The main qualitative conclusions survive, although two individual metro-gradient contrasts attenuate under clustering. The B$\to$C match-rate differences exclude zero in all four domains, at $+53.1$~pp [47.5, 58.4] for doctors, $+38.6$ [33.9, 42.8] for nursing homes, $+4.2$ [1.7, 6.7] for hospitals, and $+7.7$ [2.9, 12.2] for advisory firms. The misconduct flip holds in both directions. Arm B's pooled rate of 18.3\% [14.5, 22.6] and $>$100-employee band rate of 31.5\% [24.9, 38.4] both sit entirely above their registry bases (5.1\% and 22.6\%), while arm C's pooled 1.7\% [0.8, 2.9] and band 8.8\% [2.6, 16.8] sit entirely below. Nursing-home mean stars are 3.40 [3.21, 3.60] (arm B) and 4.44 [4.33, 4.54] (arm C), both far above the 2.88 roster mean, with non-overlapping intervals. For the metro-size gradients, arm A's doctor ($+4.6$~pp [3.1, 6.2]) and hospital ($+19.3$ [11.3, 27.3]) T1--T3 gaps and arm B's advisory-firm gap ($+16.7$ [10.3, 23.1]) exclude zero, arm B's doctor ($+6.8$ [$-1.0$, 14.4]) and hospital ($+9.9$ [$-0.6$, 21.3]) gaps are positive but marginal under clustering, and every arm-C gap includes zero. Section~\ref{sec:equity} states the gradient result with this nuance.

\paragraph{One observation per firm.}
The advisory-firm rates weight firms by how often they are recommended. Counting each matched firm once instead, arm B has 266 unique matched firms with an Item~11 rate of 9.0\% against the 5.1\% base ($z=+2.9$), arm C has 506 with 2.4\% ($z=-2.8$), and both pooled contrasts keep their direction and significance. Within the $>$100-employee band the unique-firm samples are small (65 and 44 firms) and the contrasts are directionally consistent but not individually significant (29.2\% vs.\ 22.6\%, and 13.6\% vs.\ 22.6\%). The attenuation from 18.3\% to 9.0\% in arm B shows that repeated recommendation of a few large disclosed firms carries part of the mention-weighted excess. Since a firm recommended hundreds of times is correspondingly more likely to be the one a user encounters, we regard the mention-weighted rate as the exposure-relevant estimand and the unique-firm rate as its per-provider complement.

\paragraph{Metro-expanded matching.}
The primary matcher requires the provider's registry city to be the metro's principal city, so suburban providers count as unmatched. As a complementary specification we match on name alone among providers with a practice ZIP code inside the queried CBSA, mapping registry ZIPs through the Census 2020 ZCTA-to-county relationship file and the July 2023 CBSA delineations. Any county overlap qualifies, candidates are deduplicated by NPI, CCN, or facility ID, a name with more than one candidate provider is dropped as ambiguous, and no rename-tolerant tier is applied. Match rates move as suburban recovery predicts, with doctors at 12.9\% (B) and 67.8\% (C) against 10.8\%/63.9\% primary, and nursing homes at 39.2\%/75.5\% against 32.7\%/71.2\%. Hospitals reach 36.9\%/38.9\%, above the exact-tier primary rates (33.5\%/35.8\%) and below the primary rates with the rename-tolerant subset tier (48.3\%/52.5\%), which the expanded matcher deliberately omits. The qualitative conclusions are unchanged, in that the search contrast concentrates in the same domains, arm B's tercile gradients keep their direction (doctors 17\% vs.\ 13\%, hospitals 41\% vs.\ 35\%), and arm C stays flat (doctors 66\% vs.\ 68\%, nursing homes 75\% vs.\ 80\%). Two approximations apply, since ZCTAs approximate ZIP codes and we join 2020 ZCTA geography to 2023 delineations.

\paragraph{Restaurant standardization.}
Review counts and star ratings are incommensurable scales, so Section~\ref{sec:restaurants} standardizes both against the census distribution (rating SD 0.337 stars, log$_{10}$ review-count SD 0.558). The standardized shifts are $+1.21$ SD in visibility vs.\ $+0.22$ in rating for arm A, $+1.12$ vs.\ $+0.32$ for arm B, $+0.93$ vs.\ $+0.35$ for arm C, and $+1.18$ vs.\ $+0.51$ for the Google-surfaced set, and the recommended sets' median review counts sit at the 93rd (A), 84th (B), and 82nd (C) percentiles of the census distribution. Arm A's restaurant recommendations come from a September 4, 2026 re-run of its restaurant prompts (the 55-prompt extension grid plus the ten Austin and Seattle prompts of the main grid, covering all 13 metros) on the same released gpt-oss-120b weights through a hosted API, resolved with the same browser instrument and token-match rule. We hand-classified 44 token-rule survivors as misresolutions (wrong entity, wrong city, or non-restaurant) and excluded them, and arm A's lookups postdate the arm B/C pulls by two weeks.

\section{Ethics and adverse impact}
\label{app:ethics}

This study audits software systems, not people. Every query concerns public-facing commercial services, no human participants are involved, and every registry we use is public government data published for accountability purposes. The study therefore falls outside the scope of human-subjects review.

The main ethical risk sits in the merged data rather than in the data collection. Matching recommendations to registries produces rows that pair a named provider with a misconduct disclosure or a quality rating, and published out of context such a pairing could harm the reputation of a specific doctor, facility, or firm. We therefore report aggregates only, suppress small cells, and release no table pairing an individual provider with a disclosure or rating, and the released artifacts contain only what reproduces the matching from public inputs. Two properties of the underlying measures reinforce this choice. Item~11 disclosures are records of past events rather than judgments about current conduct, and star ratings are contested instruments (Section~\ref{sec:discussion}). Our claims concern the composition of recommendation sets, never any individual provider.

Adverse impact runs in both directions. Publishing that no-search configurations fabricate referrals could erode trust in AI assistants beyond what the evidence supports; we bound the finding carefully to the audited configurations, domains, and time. Conversely, publishing which sources search relies on maps the terrain that generative engine optimization targets. We judge this acceptable because the practitioners of that industry already know it, while users and regulators mostly do not, and because the protective effects we document are contingent facts that deserve monitoring rather than trade secrets. Our query load was deliberately modest, and the audit imposed negligible cost on the services involved.

\end{document}